\documentclass{naturemodified}

\usepackage{graphicx}
\usepackage{amsmath}
\usepackage{amssymb}

\title{Entropy and Barrier-Hopping Determine Conformational Viscoelasticity in Single Biomolecules}

\author{Bhavin S. Khatri,  Masaru Kawakami,  Katherine Byrne, \\D. Alastair Smith, Tom C.B. McLeish,$^{\ast}$}

\begin{document}

\maketitle

\begin{affiliations}
\item Institute of Molecular Biophysics \& Polymer and Complex Fluids Group, School of Physics and Astronomy,University of Leeds, Leeds LS2 9JT, UK
\end{affiliations}

\begin{abstract}
Biological macromolecules have complex and non-trivial energy
landscapes, endowing them a unique conformational adaptability and
diversity in function. Hence, understanding the processes of
elasticity and dissipation at the nanoscale is important to
molecular biology and also emerging fields such as nanotechnology.
Here we analyse single molecule fluctuations in an atomic force
microscope (AFM) experiment using a generic model of biopolymer
viscoelasticity that importantly includes sources of local
`internal' conformational dissipation. Comparing two biopolymers,
dextran and cellulose, polysaccharides with and without the
well-known `chair-to-boat' transition, reveals a signature of this
simple conformational change as minima in both the elasticity and
internal friction around a characteristic force. A calculation of
two-state populations dynamics offers a simple explanation in terms
of an elasticity driven by the entropy, and friction by
barrier-controlled hopping, of populations on a landscape. The
microscopic model, allows quantitative mapping of features of the
energy landscape, revealing unexpectedly slow dynamics, suggestive
of an underlying roughness to the free energy.
\end{abstract}

\newpage

The paradigm underlying many force probe experiments is the linear
increase of tensile force on a single biomolecule with
time\cite{Evans97}. An emergent theme from constant loading rate
experiments is the propensity for conformational change in
biomolecules, from reversible processes such as chair-boat
transitions in
polysaccharides\cite{Rief97,Marszalek98,Li98,Marszalek02} and the
overstretching transition in DNA\cite{Cluzel96,Smith96,Rief99} to
the irreversible unfolding of concatamers of protein
domains\cite{Rief97+,Mitsui96,Li02,Brockwell03,Carrion-Vazquez03,Williams03}.
In addition, conformational transitions are ubiquitous in biological
processes; for example, static and dynamic changes in structure are
known to be important in many signalling processes in molecular
biology\cite{Wand01}. However despite their importance, the physical
processes that underly these transitions, in particular the role of
conformational elasticity and internal friction, are poorly
understood.

Despite the success of constant loading rate experiments, they can
provide only limited information; the elastic response function for
each molecule under reversible conditions, and at most \emph{global}
dynamical information, such as the rate of unfolding of a protein,
from irreversible stretching. A case in point is the polysaccharide
dextran, which exhibits a reversible plateau in its force-extension
response, due to a local chair-boat transition that has been shown
to be two-state in nature\cite{Rief98,Marszalek02}. Such experiments
provide the free energy difference and distance between states,
however, the dynamics of this transition are too fast for stretching
experiments to probe. A fuller understanding of the response of
single biopolymers during forced unfolding or refolding could be
provided by analysis of the \textit{local} linear viscoelastic
response. Significantly, local dissipation would give access to
finer-scale conformational dynamics, for example, the rates of
transitions between different states along the unfolding or
refolding pathways of a protein. A close analogy is found in the
macroscopic rheology of complex fluids, whose dissipative mechanical
spectra reflect dynamics of various structural, molecular and
topological transitions\cite{McLeish02}.

Recent
experiments\cite{Humphris00,Humphris02,Janovjak05,Higgins05,Okajima04,Mitsui00,
Kawakami04,Kawakami05} measuring the viscoelastic properties of
single biomolecules as a function of force, including
polysaccharides and proteins, have gone some way to achieving this
goal. The results show highly non-trivial features, particularly in
the dissipative part of the spectra, where measured frictions are
many orders of magnitude larger than that due to solvent. In
particular, the friction of dextran exhibits a minimum at a force
that coincides with the plateau in the force-extension trace,
indicating it arises through a process related to the local
conformational transitions in the chain\cite{Kawakami04,Humphris00}.
In addition, although it is clear that a plateau in the
force-extension response, should give rise to a minimum in
elasticity, the underlying statistical mechanics of this change are
not well understood\cite{Rief97,Marszalek98}. Here we seek to
understand the origins of these features in the viscoelasticity of
dextran and by doing so give broad insight to the nature of
elasticity and friction for simple conformational transitions.

Dextran and cellulose are polysaccharides that are biological
polymers composed of glucose monomers, a six-membered ring molecule,
which is known to have a number of stable
conformations\cite{Stoddart71} (Fig.1). These biopolymers differ by
the way the glucose ring is linked into the backbone of the polymer.
In dextran, one of the linkages is axial to the plane of the ring
and thus force promotes conversion from the nominally stable chair
state to a more elongated boat-like conformation, in which this
linkage is equatorial\cite{Marszalek99}, as shown in Fig.1a. This
gives rise to dextran's characteristic plateau in its force
extension response (Appendix: Fig.1). In contrast, the glucose ring
in cellulose is already near maximum elongation since all its
linkages are equatorial to the plane of the ring (Fig.1b) and
results in almost ideal freely jointed chain (FJC) properties in its
force-extension response\cite{Marszalek98, Li98} (Appendix: Fig.7).
Hence, we will show through the experimental comparison of the
viscoelasticity of these two polysaccharides, that the two-state
nature of the transition in dextran and its absence in cellulose,
provides an ideal test-bed to understand the characteristic
viscoelastic response of simple forced conformational transitions.

We determine the viscoelasticity of dextran and cellulose, using a
recently developed technique for measuring the Brownian dynamics of
single molecules under force-clamp conditions\cite{Kawakami04}.
Fig.2a summarises the experimental apparatus and procedure, with
details given in Methods. The principle of the experiment is to hold
a single molecule between tip and substrate of an AFM at constant
force, whilst observing the thermal fluctuations of the cantilever.
The fluctuations contain inherent viscoelastic information, which we
obtain via calculation of their frequency power spectral density
(PSD). A conventional proportional-integral-derivative (PID)
feedback loop with a response time of  $\sim$10 ms, monitors the
cantilever deflection signal and adjusts the piezo substrate to
maintain a constant average force ($F$) or ``force-clamp'' on the
molecule between the tip and substrate. A key idea of this technique
is that by controlling the force we probe the local viscoelasticity
of single biomolecules as they explore their energy landscape under
near equilibrium conditions. Measurement of the force-dependent
power spectra is exemplified in Fig.3 for cellulose, where it is
clear that the PSD peak position, width and amplitude are dependent
on the response of the biopolymer.

To quantify these changes and extract viscoelastic information from
the thermal spectroscopy power spectra, we model the biopolymer
using a modified Rouse model that includes local conformational
internal friction, in addition to solvent
friction\cite{Pugh75,McInnes77}. The Rouse model is a generic and
highly successful description of the coarse-grained dynamical
behaviour of polymers\cite{Rouse53,Doi}, where we note that in the
typically highly extended conformations in our experiment,
long-range hydrodynamics \cite{Zimm56} give only logarithmic
corrections to local drag. The Rouse with Internal Friction (RIF)
polymer is represented as a series of beads with solvent friction
$\zeta_{s0}$, connected by spring and dashpots of elasticity
$\kappa_0$ and internal friction $\zeta_{i0}$. In the continuum
limit, internal friction adds an extra term in the standard Rouse
equation, which describes a dissipative force proportional to the
rate of change of local conformation, represented as the
coarse-grained curvature of the chain,

\begin{equation}\label{RIFeqn}
\zeta_{s0}\frac{\partial\boldsymbol{R}(n,t)}{\partial t}
=\biggl(\kappa_0 + \zeta_{i0}\frac{\partial}{\partial
t}\biggr)\frac{\partial^2\boldsymbol{R}(n,t)}{\partial n^2} +
\boldsymbol{f}(n,t),
\end{equation}

\noindent where $\boldsymbol{R}(n,t)$ represents the space curve of
the polymer with contour variable $n$, subject to a local Langevin
force $\boldsymbol{f}(n,t)$, which is uncorrelated for different
times. Normal mode solutions of this equation decay in a single
exponential manner with a mode dependent relaxation time, $\tau_p =
\frac{N\zeta_p}{2\pi^2\kappa_0 p^2}$, where $p$ is the mode number.
the effective mode friction is renormalised compared to standard
Rouse theory to $\zeta_p = 2N\zeta_{s0} +
\frac{2\pi^2p^2\zeta_{i0}}{N}$, where intuitively, the new term
accounts for an increasing internal friction of higher curvature
short wavelength modes. AFM experiments probe the end-to-end vector
of the polymer, whose response can be found by summing over all odd
modes; in frequency space this gives the following useful closed
form expression for the dynamic compliance:

\begin{equation}\label{GreensRIF}
J_{\Delta R}(\omega)
=\frac{2N}{\pi\kappa_0}\frac{\tanh{\biggl(\frac{\pi}{2}\sqrt{\frac{i\omega\tau_R}
{1+i\omega\tau_i} }\biggr)}}{\sqrt{i\omega\tau_R(1+i\omega\tau_i)}},
\end{equation}

\noindent where $\tau_R = N^2\zeta_{s0}/\pi^2\kappa_0$ is the
contribution to the relaxation time of the first mode due to solvent
friction and $\tau_i=\zeta_{i0}/\kappa_0$ is the mode-independent
contribution to the relaxation time due to internal friction. This
model successfully encompasses the behaviour of both types of
friction; in the limit of large internal friction
($\tau_R<<\tau_i$), Eq.(\ref{GreensRIF}) reduces to a single mode
spring and dashpot model, $J_{\Delta
R}(\omega)=\frac{N}{\kappa_0}\frac{1}{(1+i\omega\tau_i)}$ and when
solvent friction dominates ($\tau_R>> \tau_i$) to the Rouse model,
given by the limiting form, $J_{\Delta
R}(\omega)=\frac{2N}{\pi\kappa_0}
\tanh{\left(\frac{\pi}{2}\sqrt{i\omega\tau_R}\right)}/\sqrt{i\omega\tau_R}$,
till a critical frequency $1/\tau_i$, when the internal friction of
high curvature modes dominates to give single mode relaxation again.
The Fluctuation-Dissipation Theorem (FDT)\cite{ChaikenLubensky},
$P(\omega) =- 2k_BTJ''(\omega)/\omega$ is then used to calculate the
total power spectrum $P(\omega)$ of a RIF polymer combined with a
SHO response of the cantilever and cantilever, where $J''(\omega)$
is the imaginary part of the response function $J(\omega)$.

Shown in Fig.4a are the effective monomer elasticity of cellulose
and dextran, from the RIF model fits and normalised by contour
length. In previous work\cite{Kawakami04} we showed that calculating
the elasticity spectrum directly from the numerical derivative of
extensible FJC fits, and secondly from the thermal spectroscopy
method agree very well. We verify that using the more refined RIF +
cantilever model to analyse the PSD, also provides very good
agreement. As previous studies have
shown\cite{Rief97,Marszalek02,Marszalek98,Li98,Humphris00,Humphris02,
Kawakami04, Kawakami05}, at low force (in these experiments),
elasticity is due to the reduction of chain conformational entropy
as it approaches its contour length, after which contour length
elongation with constant elasticity becomes more favourable. At
higher force, however, the minimum in the elasticity spectra for
dextran at $\sim 1000$pN, which is absent in the cellulose spectrum,
marks a clear signal of the conformational transition in the former.

The key advance afforded by using the RIF model in analysing the PSD
is the new information about the two sources of dissipation, not
distinguished in previous work\cite{Humphris00,Humphris02,
Kawakami04, Kawakami05}; the solvent friction and internal friction
of the single biomolecule. We find consistently from the RIF
analysis, that solvent friction is very small within the errors of
this experiment ($\leq$ 0.01$ \mu$gkHz). Hence, these chains are
`short', as defined by $N\ll\sqrt{\zeta_{i0}/\zeta_{s0}}$
\cite{deGennes}, where $N\sim$ 400, which indicates that dissipation
is dominated by internal friction at high stretch, and explains the
success of the spring and dashpot model in previous modelling of the
dissipation of dextran \cite{Humphris00,Humphris02, Kawakami04,
Kawakami05}. The internal friction force spectrum itself exhibits
non-trivial behaviour as shown by the comparison of cellulose and
dextran in Fig.4b. At low force, both polymers show an increasing
internal friction with force, followed by a plateau. Crucially, at
higher forces, the spectra of cellulose and dextran differ;
qualitatively, the minimum in the internal friction force spectrum
of dextran at $\sim 1000$pN and its absence in cellulose, confirms
that source of this change in the friction of dextran is from the
chair-boat conformational transition of the glucose ring.

To make this conclusion more concrete we link the features of the
experimental elasticity and friction force spectra to the
conformational transition in dextran, using a simple model of
population dynamics on a discrete 2-state energy landscape, which we
show predicts the same viscoelastic signature of simple forced
transitions, as seen in Fig.4. The parameters of the discrete
2-state model are as described in Fig.5a, in which we assume
populations obeys Boltzmann statistics and dynamics follow activated
Arhennius transition rates. Using an approach similar
to\cite{McNamara89,Braun04+}, the effective response of the
populations at a frequency $\omega$ can be calculated by applying an
oscillatory force $f_0\cos{\omega t}$ to the energy landscape. The
dynamics of the population $p(t)$ in state 1 (say, the short state,
so that probability of extended state is $1-p(t)$) are then
described by $\frac{dp}{dt} = -(\lambda_{12}(t)+\lambda_{21}(t))p(t)
+\lambda_{21}(t)$, where the rates $\lambda_{12},\ \lambda_{21}$
vary with time due to the oscillating perturbation of the landscape.
In the Brownian linear response limit ($f_0x\ll k_BT$), there are
in-phase and out-of-phase oscillating solutions to this differential
equation, such that the extensional response of the monomer is a
simple spring and dashpot
$J_{12}(\omega)=\frac{1}{\kappa_{12}+i\zeta_{12}\omega}$, for which
we identify the effective elasticity and friction as:

\begin{equation}\label{KappaHop}
\kappa_{12}(F) = \frac{k_BT}{(\Delta
x)^2}\frac{1}{p_0(F)(1-p_0(F))},
\end{equation}
%

\begin{equation}\label{ZetaHop}
\zeta_{12}(F) = \frac{k_BT}{(\Delta x)^2}(\tau_{12}(F)
+\tau_{21}(F)),
\end{equation}

\noindent and where $p_0(F)=(1+e^{-\beta\Delta G(F)})^{-1}$ is the
equilibrium Boltzmann probability for the short state. In addition,
$\tau_{12}$ is the forward hopping time between states and
$\tau_{21}$ is the corresponding backward time, where in general
$\tau_{ij}=\tau_{0}e^{\beta\Delta G^{\ddagger}_{ij}(F)}$, with
$\beta=1/k_BT$, $\Delta G^{\ddagger}_{ij}$ the free energy barriers
for interconversion and $\tau_0=2\pi\zeta_b/\kappa_b$ is a prefactor
that arises from mapping the Kramers' first passage problem on a
continuous free energy landscape\cite{Kramers40} $G(x)$ to a
discrete description (Fig.5a), where $\zeta_b$ and $\kappa_b$ are
the effective friction and curvature of the barrier.

Plotting these (Fig.5b\&c - on a natural logarithmic scale to
emphasise their exponential nature) we see a characteristic minimum
in both the elasticity and internal friction force spectra. In the
former case, it is clear that the source of the elasticity is
\textit{entropic} in nature and not enthalpic as has been previously
asserted\cite{Marszalek98}; force controls the shape of the energy
landscape or the relative populations of monomers in short or
extended states and hence, the effective `size of box' that the
monomer can explore. So Eq.(\ref{KappaHop}) is an expression of the
equipartition theorem $\kappa = k_BT/\langle\Delta b^2\rangle$,
where $\langle\Delta b^2\rangle$ is the mean square fluctuations of
the monomer; in Fig.5b at low force, $\Delta G(F)$ is large and
positive, hence monomers are confined to the short state,
fluctuations $\langle\Delta b^2\rangle$ are small and the effective
stiffness is large. As force decreases the energy difference,
populations spread across the two states and the effective size of
the box $\langle\Delta b^2\rangle$ increases, causing the stiffness
to decrease (exponentially). The stiffness subsequently passes
through a minimum when $\Delta G(F)=0$ and $\langle\Delta
b^2\rangle$ is maximum, corresponding to a state of maximum entropy,
when the probabilities to be in either of the states are equal. On
further increase of force, $\Delta G(F)$ becomes negative and
monomers become increasingly confined to the extended state
($\langle\Delta b^2\rangle$ decreasing) and the stiffness increases
exponentially. It is simple to see that the elasticity is purely
entropic, since any enthalpic contributions to the free energy
difference $\Delta G_0$ can contribute only linearly to the free
energy as the extension of the monomer is increased. Thus, the
molecular elasticity of a monomer is defined by its entropy on a
discrete energy landscape.

Eq.(\ref{ZetaHop}) predicts that the internal friction for a 2-state
landscape is proportional to the sum of the times to interconvert
from state 1 to 2 and back, from state 2 to 1. Applying a force to
the monomers changes the activation barriers to interconversion,
which changes the average time to interconvert and thus ultimately,
the internal friction. Fig.5c shows schematically how the internal
friction should vary with force in a discrete two-state landscape.
As force lowers the barrier $\Delta G^\ddagger_{12}(F)$ of
interconverting from 1 $\rightarrow$ 2, the internal friction should
decrease, passing through a minimum when the barriers on either side
are approximately equal (when $x_1 = x_2$ this occurs at exactly
$\Delta G^\ddagger_{12} = \Delta G^\ddagger_{21}$) and then increase
again at high force as the barrier for the reverse transition
($\Delta G^\ddagger_{21}$) and hence $\tau_{21}$, becomes large. It
is interesting to note that, whilst the hopping time passes through
a minimum, the corresponding relaxation time
$\tau=\tau_{12}^{-1}+\tau_{21}^{-1}$ must pass through a maximum,
since relaxation is dominated by the \textit{smallest} barrier.
Hence, on average fluctuations away from equilibrium occur on a
hopping timescale $\tau^*\sim\tau_{12}+\tau_{21}$, whilst relaxation
back to equilibrium occurs on the timescale $\tau$. We see
Eq.(\ref{ZetaHop}), is a microscopic fluctuation-dissipation
relation for a discrete bistable landscape, that links friction to
the timescale for fluctuations due to activated barrier-hopping.

Useful information about the position of the transition state can
also be obtained by analysing the relative positions of the minima
in the elasticity and internal friction spectra. The difference in
the forces at which the minima occur $\Delta F$, can be found from
the derivatives of Eq.(\ref{KappaHop}) and Eq.(\ref{ZetaHop}):

\begin{equation}\label{DeltaF}
\Delta F=\frac{k_BT}{\Delta x}\ln{\biggl(\frac{x_1}{x_2}\biggr)},
\end{equation}

\noindent and thus provides information on the relative position of
the transition state, $x_1$ or $x_2$ ($\Delta x=x_1+x_2$).

The results of this population dynamics model, thus provide a simple
way to understand the minima in the elasticity and internal friction
spectra, in terms of entropy and barrier-hopping. However, to
understand the entire force regime ($\sim 100\rightarrow 1500$pN),
in addition to the viscoelasticity of the 2-state conformational
transition, we need to include the physics of the chain at low and
intermediate forces, before the critical force at which the
conformational transition occurs. At low force we use a Frictional
Freely Jointed Chain (FFJC) model (Appendix) of rods interconnected
with joints with constant friction $\zeta_{\theta}$ to give an
elasticity $\kappa_{FJC}(F) = \frac{F^2}{k_BT}$ and an internal
friction that increases linearly with force, $\zeta_{FJC}(F) =
\frac{\zeta_{\theta}}{2k_bTb}F$, which are both valid at high
stretch ($F\gg k_BT/b\sim 4$pN for $b\sim1$nm). At intermediate
force we account for the local viscoelasticity of stretching a
dextran monomer in the short or extended states, using constant
elasticities $\kappa_1$, $\kappa_2$, and internal frictions
$\zeta_1$, $\zeta_2$, respectively. We assume that these processes
add mechanically in series, since they provide independent and
additive extensions to the overall chain length (see Methods).


%


Fitting to the elasticity force spectra of cellulose and dextran
(normalised by contour length), we find excellent agreement as shown
in Fig.4a, where the solid line represents the full elasticity
Eq.(\ref{ElasticitySum}) generated using the average of the
parameters determined over a number of single molecule experiments
(Cellulose: $\kappa_1=36000\pm18000$ pN/nm, $b=1\pm0.5$nm; Dextran:
$\Delta G_0=16.5\pm0.4 k_BT$, $\Delta x=0.066\pm0.005$nm,
$\kappa_1=10000\pm1000$pN/nm, $\kappa_2=39000\pm2000$pN/nm,
$b=0.63\pm0.02$nm). These values agree well with the literature
\cite{Janshoff00, Marszalek02, Marszalek98, Rief97}. With confidence
we can describe the whole elastic force spectra for both cellulose
and dextran; at low force (below 800pN) stiffness increases as
entropy is lost due to the orientation of monomers along the line of
force and finally reaches a plateau representing a constant
stiffness due to the enthalpy of stretching the bonds comprising the
glucose ring. However, the response of dextran differs dramatically
at higher force as the more extended state becomes thermodynamically
favourable. Within the framework of the 2-state model presented, the
subsequent decrease in stiffness can be understood since it becomes
more \textit{entropically} favourable for the chain to elongate.
Interestingly, in the elasticity spectrum of dextran, at around
400-500pN, the model slightly, but consistently, underpredicts the
elasticity below this force and overpredicts it above this force.
This plateau may be explained by the entropic elasticity of other
internal states, possibly the C5-C6 bond rotation in dextran
\cite{Lee04,Rief97}.

%
In performing fits to the internal friction spectra, all elastic
parameters are constrained to values obtained from fits to the
elasticity spectra (see Methods). Below we discuss quantitative
values of each of these friction processes separately, even though
actual fits were performed globally across the whole force range.

Firstly, we examine the effective internal friction associated with
stretching the glucose monomers in their various conformations. For
cellulose, we find $\zeta_1=110\pm50 \mu$gkHz and for dextran,
$\zeta_1=25\pm10\mu$gkHz and $\zeta_2=120\pm50\mu$gkHz, for the
short and extended states, respectively. Strikingly, these numbers
are roughly 7 orders of magnitude larger than the friction expected
due to solvent ($\zeta=6\pi\eta b\sim 10^{-5}\mu$gkHz for
$b\sim1$nm). The most plausible source for such a high local
effective friction is roughness in the free energy landscape. A
model of dynamics on a rough Gaussian landscape with RMS energy
fluctuations $\varepsilon$\cite{Zwanzig88} predicts a sensitive
enhancement to the effective friction constant $\zeta^* =
\zeta\exp{(\varepsilon/k_BT)^2}$, giving an effective roughness for
stretching cellulose and dextran as $\varepsilon\approx 4k_BT$. For
comparison, recent constant loading rate experiments\cite{Nevo05} on
the protein imp-$\beta$, using theoretical results
in\cite{Thirumalai03} suggest a Gaussian roughness of order
$\varepsilon\approx 5.7 k_BT$. In the case of these polysaccharides,
this roughness may arise from the many sub-states separated by
barriers that must be traversed in stretching the monomers; for
example; for example, there are many conformations of a glucose
ring, (in total 14 canonical chair, boat and twist-boat
conformations, separated by 12 half-chair and 12 envelope
conformational transition states\cite{Stoddart71,Ionescu05}), which
will contribute to extension and may become more or less favourable
under tension. In addition, the hydroxyl groups of glucose give rise
to the possibility of intra and intermonomer hydrogen bonding, as
well differing degrees of solvent accessibility. Such states are
particularly suggested by slow undulations in the elasticity and
internal friction spectra of cellulose for forces greater than
$1000$ pN.

At low force, we see a similar picture for the `joint' friction of
the FFJC model, obtaining values of the order
$\zeta_{\theta}\sim1\mu$gnm$^2$kHz (cellulose:
$\zeta_{\theta}=0.9\pm0.7 \mu$gnm$^2$kHz; whilst for dextran errors
from fits suggest $\zeta_{\theta}<1.2 \mu$gnm$^2$kHz). These numbers
are roughly 6 orders magnitude greater than the friction of a rod of
length $b$ rotating in a solvent ($\pi\eta b^3/4\sim 10^{-6}
\mu$gnm$^2$kHz). We can again appeal to an underlying molecular
explanation, where joint friction is due to hopping between dihedral
angular states, with an average hopping time of
$\tau_{hop}\sim\zeta_{\theta}/k_BT\approx 0.25$msec. Again, these
very slow dynamics are suggestive of an underlying roughness to the
rotational free energy ($\varepsilon\approx 3.7 k_BT$, where
$\zeta^*/\zeta\sim10^6$).

In the case of dextran, the marked decrease in internal friction
around $\sim 1000$pN, contains information on the dynamics of
interconversion between the short and extended states, for which
Eq.(\ref{ZetaHop}) provides a simple model. In principle, fitting to
the internal friction spectra would determine the position of the
barrier $x_1$ (with constraint $x_2 = \Delta x-x_1$) and the
zero-force interconversion times $\tau_{12}(F=0)$ and
$\tau_{21}(F=0)$ (measurements at different temperatures could in
principle, determine the free energy barriers for conversion in each
direction). However, although we find very good fits around the
transition region, they are underdetermined, due to the low
frequency restriction of the data.

To constrain our fits further, we use Eq.(\ref{DeltaF}). Inspecting
Fig.4a\&b (solid squares), indicates that $\Delta F\approx
0\pm100$pN, given that the spacing of points in the spectra is
approximately 100pN. However, negative values of $\Delta F$ imply
from Eq.(\ref{DeltaF}) a transition state that is closer to the
short state than long ($\frac{x_1}{x_2}< 1$), which is not feasible
on geometric grounds, given that its curvature is roughly four times
smaller than the extended state ($\frac{\kappa_1}{\kappa_2}\approx
\frac{1}{4}$) and that the forward free energy barrier must obey
$\Delta G^{\ddagger}_{012}> \Delta G_0 (= 16.5k_BT)$. Hence, a
reasonable assumption is that $0<\Delta F<100$pN, implying a
position of the transition state in the region $0.033<x_1<0.053$nm.
Fitting to the friction spectra, so as to satisfy this constraint on
$x_1$, we find $1$ns$<\tau_{21}(0)<100$ns and $0.01$
s$<\tau_{12}(0)<1$ s (see Methods). From Kramers' theory
\cite{Kramers40} of activated diffusive barrier crossing, the
exponential prefactors of these times are related to the curvature
$\kappa_b$ and friction $\zeta_b$ of the barrier, which when mapped
onto a discrete landscape is given by $\tau_0=2\pi\zeta_b/\kappa_b$.
Thus, given an order of magnitude estimate of the barrier friction
$\zeta_b\sim\frac{1}{2}(\zeta_1+\zeta_2)\approx70\mu$gkHz and that
$\tau_0<\tau_{21}(0)$, we find that the barrier must be very sharp;
given by the following approximate bound, $\kappa_b> 10^6$pN/nm.
Fig.6 shows a graphical to-scale reconstruction of the free energy
landscape based on the parameters extracted from the modelling of
the viscoelastic force spectra of dextran.

In summary, we have shown how macroscopic ideas of elasticity and
friction can be extended to the study of the energy landscape of
conformational transitions. Eq.\ref{KappaHop} and Eq.\ref{ZetaHop}
are in essence microscopic equivalents of the equipartition theorem
and the diffusive fluctuation-dissipation relation, where the
spatial and temporal properties of the fluctuations are determined
by the shape of the energy landscape, which in turn determine its
effective elasticity and friction. In the case of dextran, applying
tension to its energy landscape, drives an entropic transition,
where elasticity and friction decreases as the populations become
more spread and barriers are lowered. These ideas are of wide
relevance; from applications to the field of molecular
nanotechnology, where microscopic processes of elasticity and
internal friction may guide and constrain engineering design, to
understanding fundamental processes of molecular biology, such as
the study of internal transitions in biomolecules, including the
action of molecular motors, allosteric signalling, force-sensing
between cells, stretching transitions in DNA and RNA, and emerging
data on elasticity and dissipation from the fluctuations of a
refolding protein.


\bibliography{Arxiv}

\bibliographystyle{naturemag}

\begin{methods}

\subsection{Experimental Materials \& Methods}

The protocol used for thermal force-clamp spectroscopy is as
described in\cite{Kawakami04}, we summarise the procedure here. The
first part of the experiment follows conventional force-spectroscopy
protocol, where the cantilever is pressed into a polysaccharide
monolayer with a force $\sim10$ nN for $\sim1$ s, after which it is
retracted from the substrate at a constant speed. When a
pre-determined force set-point is reached, the force-clamp protocol
is initiated, which involves either reducing force in discrete steps
of $\sim100$ pN and being held for $\sim3$ s, or reducing force
slowly and continuously at $\sim8$ pN/s. In some measurements we
have used this latter continuous approach, however, both procedures
produce the same results within the errors of each method (not
shown). In either method the force is controlled using a
proportional-integral-derivative (PID) feedback loop with a response
time of $\sim10$ ms, whereby the cantilever substrate separation is
adjusted to maintain a certain cantilever deflection. A response
time of $\sim 10$ ms means the feedback loop cannot respond to
fluctuations faster than $10$ ms. Thus, for frequencies greater than
$\sim0.1$ kHz, an average force $F$ is maintained. After the
force-clamp phase, the cantilever is again retracted from the
substrate at a constant speed, till at some critical force the
polymer detaches. Immediately after detachment, the PSD of the free
cantilever is recorded as the cantilever is brought towards the
substrate in 30 nm steps. These free cantilever PSD are then fit
using a simple harmonic oscillator model (SHO), $P_c(\omega) =
\frac{2k_BT\zeta_c}{(\kappa_c- m_c\omega^2)^2 +\zeta_c^2\omega^2}$,
obtaining the cantilever effective stiffness $\kappa_c$, friction
constant $\zeta_c$ and mass $m_c$. These parameters then serve as
constraints in the curve fits to the power spectra of the
cantilever/molecule system.

To extract the elasticity, internal and solvent friction as
functions of force, we treat the force clamp experiment as two
linear system elements in parallel, since the change in extension of
the polymer and cantilever are the same at their point of contact.
It can be shown (Appendix), that for a system in parallel the total
dynamic compliance of the system $J_T(\omega)$ is given by

\begin{equation}\label{ParAdd}
J_T(\omega)= \frac{J_X(\omega)J_{\Delta
R}(\omega)}{J_X(\omega)+J_{\Delta R}(\omega)},
\end{equation}

\noindent where $J_X(\omega)$ is the dynamic compliance of the
cantilever, for which we use a SHO model
($J_X(\omega)=(\kappa-m\omega^2+i\zeta\omega)^{-1}$). This is just
the frequency-dependent extension of the parallel addition that
arises naturally in our experiment. We then use the fluctuation
dissipation theorem (FDT) $P(\omega) =- 2k_BTJ''(\omega)/\omega$
\cite{ChaikenLubensky} ($J''$ represents the imaginary part of the
complex function $J$) and Eq.(\ref{GreensRIF}) \& Eq.(\ref{ParAdd})
to calculate the total power spectrum of the cantilever + RIF
polymer. This enables measurement of the elasticity, internal
friction and solvent friction as functions of force, for example, as
shown by the RIF model fits to the power spectra of cellulose in
Fig.3. In fitting the PSDs to the RIF+cantilever model, we constrain
the chain solvent friction to be between $0<\zeta_s<0.1$ $\mu$gkHz,
since a reasonable estimate of the solvent friction is given by
$6\pi\eta L_c approx 4\times10^{-3}$ $\mu$gkHz, where the contour
length $L_c \approx 200$ nm, is the typical contour length of
molecules in the experiment, representing the maximum effective
hydrodynamic radius of the chain\cite{deGennes}.

\subsection{Fitting to Elasticity and Internal Friction.}

In both the elasticity and internal friction force spectra, there
are a number of different physical processes that underly the
observed behaviour. A reasonable assumption is that the noise on
each physical processes is uncorrelated, so that the total power
spectrum is the sum of the power spectra of each process. In the low
frequency regime of the experiments ($\omega\tau\ll1$), the rules
for summing the elasticity and frictions of the different processes
are then:

\begin{equation}\label{ElasticitySum}
\kappa(F)=\left(\kappa_{FJC}^{-1}(F)+\kappa_{12}^{-1}(F)+
    p_0(F)\kappa_1^{-1}+(1-p_0(F))\kappa_2^{-1}\right)^{-1},
\end{equation}

\begin{equation}\label{FrictionSum}
\zeta(F)=\kappa^2(F)\biggl(\frac{\zeta_{FJC}(F)}{\kappa_{FJC}^2(F)}+
    \frac{\zeta_{12}(F)}{\kappa_{12}^2(F)}+p_0(F)\frac{\zeta_1}{\kappa_1^2}+
    (1-p_0(F))\frac{\zeta_2}{\kappa_2^2}\biggr),
\end{equation}

\noindent where importantly, the bond elasticities of the short and
extended states are weighted by the probability to be in those
states at a given force, where $p_0(F)=(1+e^{-\beta\Delta
G(F)})^{-1}$. In fitting to the internal friction force spectra, we
use the parameters extracted from fitting to the elasticity spectra
as a constraint to the fits. Through trial and error with fits with
different fixed values of $\tau_{21}(0)$, we found the values of the
zero-force backward hopping time that correspond to the bound
calculated on $x_1$ in the main text. We checked that the values of
$\tau_{12}(0)$ also determined from the fits, were consistent with
detailed balance ($\tau_{12}(0)/\tau_{21}(0)=e^{\beta\Delta G_0}$).
\end{methods}

\begin{addendum}

 \item We thank EPSRC (UK) for financial support. M.K. was a JSPS Visiting
Research Fellow and is now supported by the EPSRC. T.C.B.M. is an
EPSRC Advanced Fellow. We thank Igor Neelov and Peter Olmsted,
School of Physics \& Astronomy, University of Leeds and Stuart
Warriner, School of Chemistry, University of Leeds, for fruitful
discussions. We are particularly grateful to Sheena Radford, Astbury
Centre for Structural Molecular Biology, University of Leeds, for
many useful and stimulating discussions.

Correspondence and requests for materials should be addressed to
T.C.B.M.(email:t.c.b.mcleish@leeds.ac.uk).

\end{addendum}

\noindent {\bf Figure. 1. Structure of dextran and cellulose.} (a)
Simplified diagram of the molecular structure of dextran, which is
an $\alpha$-(1$\rightarrow$6) linked polysaccharide of glucose,
where the monomer length is defined by the distance between adjacent
non-ring oxygens on the backbone, as shown schematically. The
$\alpha$ linkage at C1 is axial in the lowest energy $^4C_1$
\textit{chair} conformation\cite{Stoddart71} (above), which under
application of a tensile force-field promotes one of a number of
more elongated \textit{boat} or \textit{skew-boat} conformations, of
which the $^{1,4}B$ is shown
(below)\cite{Marszalek98,Stoddart71,ODonoghue00,Ionescu05}. The
increased length $\Delta x$ gives rise to a plateau in
force-extension measurements (Appendix Figure 7) as the more
elongated boat-like conformations are populated under increasing
force. (b) Cellulose on the other hand is a
$\beta$-(1$\rightarrow$4) linked polysaccharide of glucose, whose
equatorial linkage at C1 in the chair state, means the monomer is
already near maximum elongation and a its force-extension behaviour
follows simple polymer elasticity models due to reduction of chain
entropy at high stretch (Appendix Figure 7).

\noindent {\bf Figure. 2. Force-clamp thermal noise spectroscopy.}
(a) Schematic diagram of the experimental setup for thermal noise
spectroscopy. (b)\&(c) show the force clamp protocol used: (b)
typical experimental force-extension traces showing a retract,
approach and retract cycle, using dextran as the sample polymer,
where the traces have been offset for clarity. The characteristic
shape of the final curve (curve 3), which in the case of dextran
exhibits a shoulder indicative of the well-known conformational
transition in dextran\cite{Rief97}, confirms that only a single
molecule was attached. (c) experimental force-time trace, where the
numbers and colours correspond to the same sequence in (b). The
force-clamp phase (phase 2) lasts for a total of 42 seconds, where
the dextran polymer is held for 3 seconds at each of 14 discrete
forces (the initial and final extensions, phases 1 \& 3, are shown
on an expanded time scale).

\noindent {\bf Figure 3. Force dependent PSD of single molecule of
cellulose.} Comparison of the PSD of fluctuations of cantilever tip,
when free (black circles) and with a single cellulose molecule
attached, held with forces of 320 pN (green), 620 pN (purple) and
920 pN (red). The solid lines correspond to fits using either a
simple harmonic oscillator model for the cantilever (black), or the
RIF model of the biopolymer combined with the cantilever (green,
purple and red solid lines) described by Eq.(\ref{GreensRIF}) \&
Eq.(\ref{ParAdd}).

\noindent {\bf Figure 4. Viscoelastic force spectrum of cellulose
and dextran.}  (a) Elasticity force spectrum and (b) internal
friction force spectrum multiplied by contour length of each
molecule $L_c$ (giving the inverses of the compliance and mobility
per unit length) for cellulose (solid diamonds) and dextran (solid
squares, where $L_c$ is obtained from FJC fits to their respective
force-extension traces (Appendix Figure 7)). Data points represent
measurements using thermal force clamp spectroscopy, where different
colours represent separate single molecules. The solid lines
represent curves generated using the full elasticity (a) and
internal friction (b) expressions given in Eq.(\ref{ElasticitySum})
\& Eq.(\ref{FrictionSum}), using the average of the parameters
determined over all of single molecule experiments (see main text),
apart from $\zeta_{\theta}=0.6\mu$gnm$^2$kHz (half the upper bound
in the main text) and 2-state internal friction parameters derived
in the text consistent with a fixed zero-force backward
interconversion time $\tau_{21}(0)=100$ns (i.e. $\Delta
x_1=0.053$nm, $\tau_{12}(0)=1$s). Horizontal error bars represent an
approximate 10\% systematic error between experiments in determining
the true force scale, through errors in measuring cantilever
elasticity and cantilever deflection sensitivity. Vertical error
bars represent errors from the fits to the PSDs.

\noindent {\bf Figure 5. Viscoelastic Force Spectrum on a discrete
bistable landscape.} (a) Schematic diagram of the discrete free
energy landscape used to calculate the elasticity and internal
friction force spectra (Eq.(\ref{KappaHop}) \& Eq.(\ref{ZetaHop})).
(b) Elasticity force spectrum on a discrete 2-state landscape; force
controls free energy difference $\Delta G(F)=\Delta G_0-F\Delta
x_{12}$ and hence spread $\langle\Delta b^2\rangle$ and elasticity
$\kappa_{12}(F)=k_BT/\langle\Delta b^2\rangle$). Elasticity is
entropic in nature as elasticity decreases in direction of
increasing entropy of monomers. (c) Internal friction force spectrum
for a discrete 2-state landscape; force controls activation barrier
heights ($\Delta G^{\dagger}_{12}(F)=\Delta
G_{012}^{\ddagger}-F\Delta x_1$, $\Delta G^{\dagger}_{21}(F)=\Delta
G_{021}^{\ddagger}+F\Delta x_2$), and therefore also the internal
friction. Hence, at a given force, internal friction is dominated by
the activation barrier that is largest.

\noindent {\bf Figure 6. To-scale reconstruction of continuous free
energy landscape of glucose,} based on parameters extracted from
theoretical modelling of the viscoelastic force spectra of dextran.
Dashed features indicate areas of landscape that are uncertain, for
example position of barrier, or information unattainable with
current experiments like the activation barrier heights. Barrier
curvature shown is $\kappa_b=10^6$pN/nm. Grey lines indicate a
roughness to the landscape with RMS deviation
$\varepsilon\approx4k_BT$, as a plausible interpretation for
significantly enhanced friction of wells. $\Delta G'_0=\Delta G_0 +
k_BT\ln(\sqrt{\kappa_1/\kappa_2})=(16.5-\ln2)k_BT\approx 15.8k_BT$
is the free energy difference between the minima of a continuous
landscape, which excludes the entropy of vibrations of the wells.

\begin{center}
\includegraphics[width=8cm]{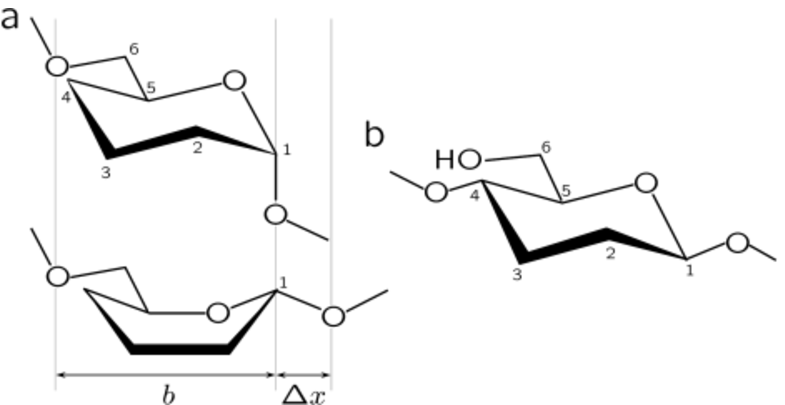}
\end{center}
\begin{center}Figure 1\end{center}

\begin{center}
\includegraphics[width=14cm]{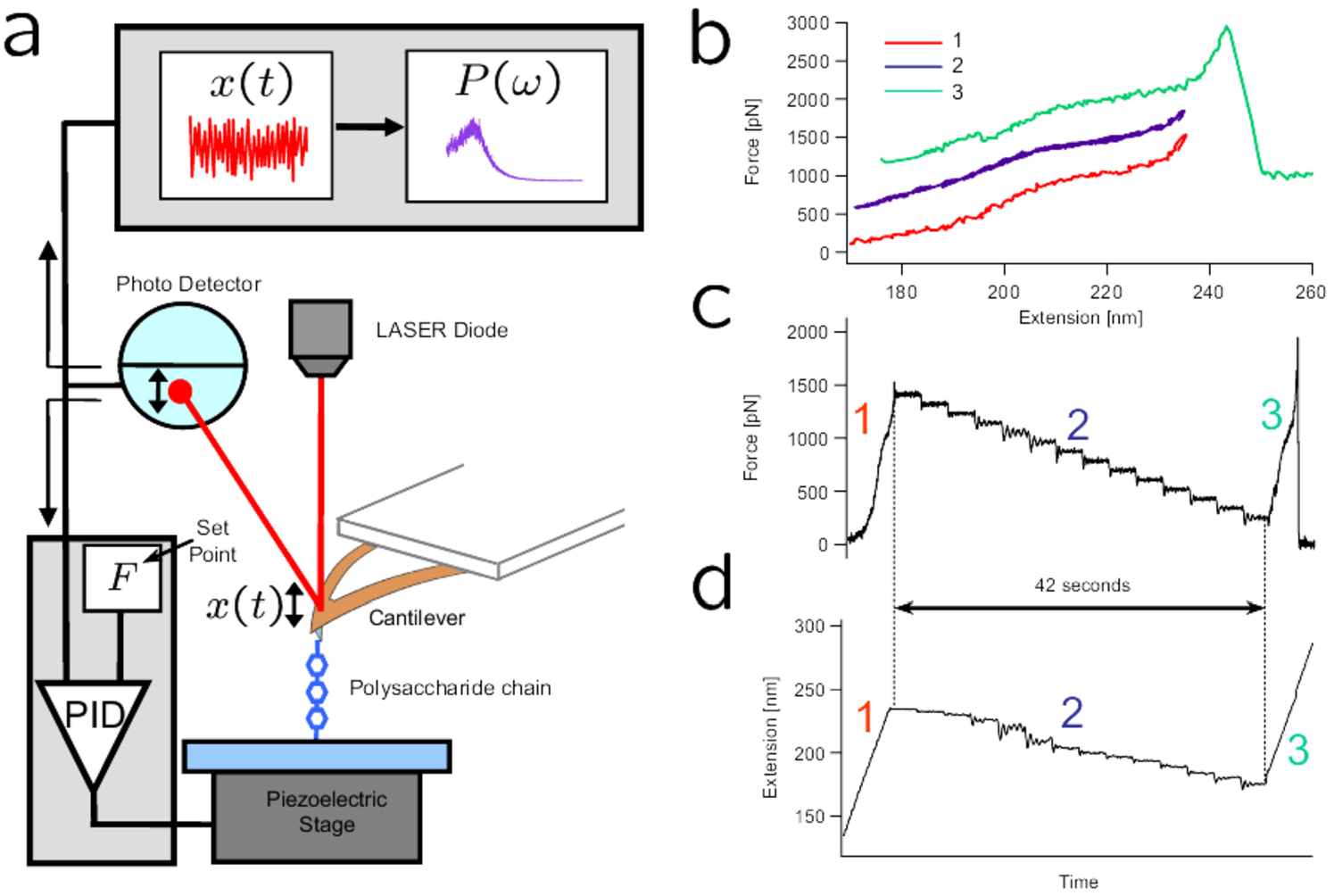}
\end{center}
\begin{center}Figure 2\end{center}

\begin{center}
\includegraphics[width=8cm]{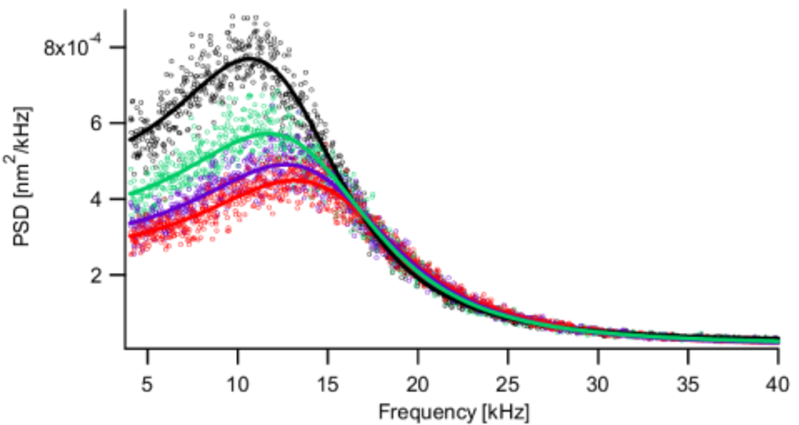}
\end{center}
\begin{center}Figure 3\end{center}

\begin{center}
\includegraphics[width=10cm]{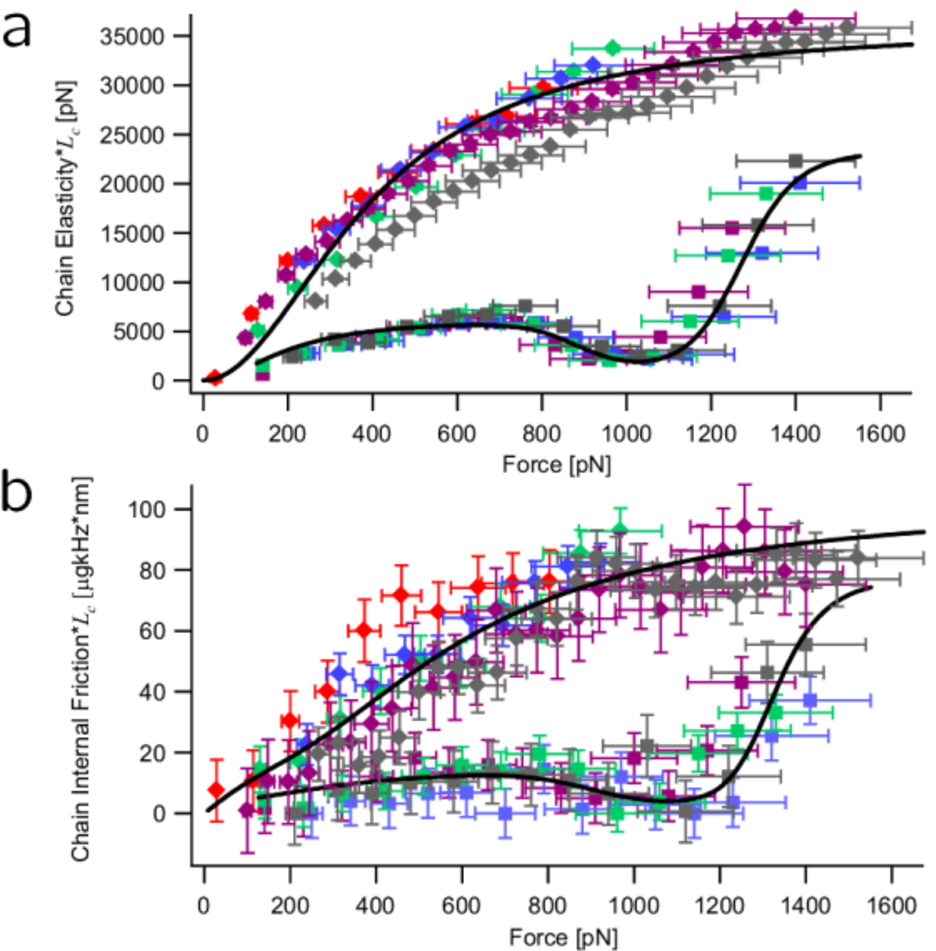}
\end{center}
\begin{center}Figure 4\end{center}

\begin{center}
\includegraphics[width=8cm]{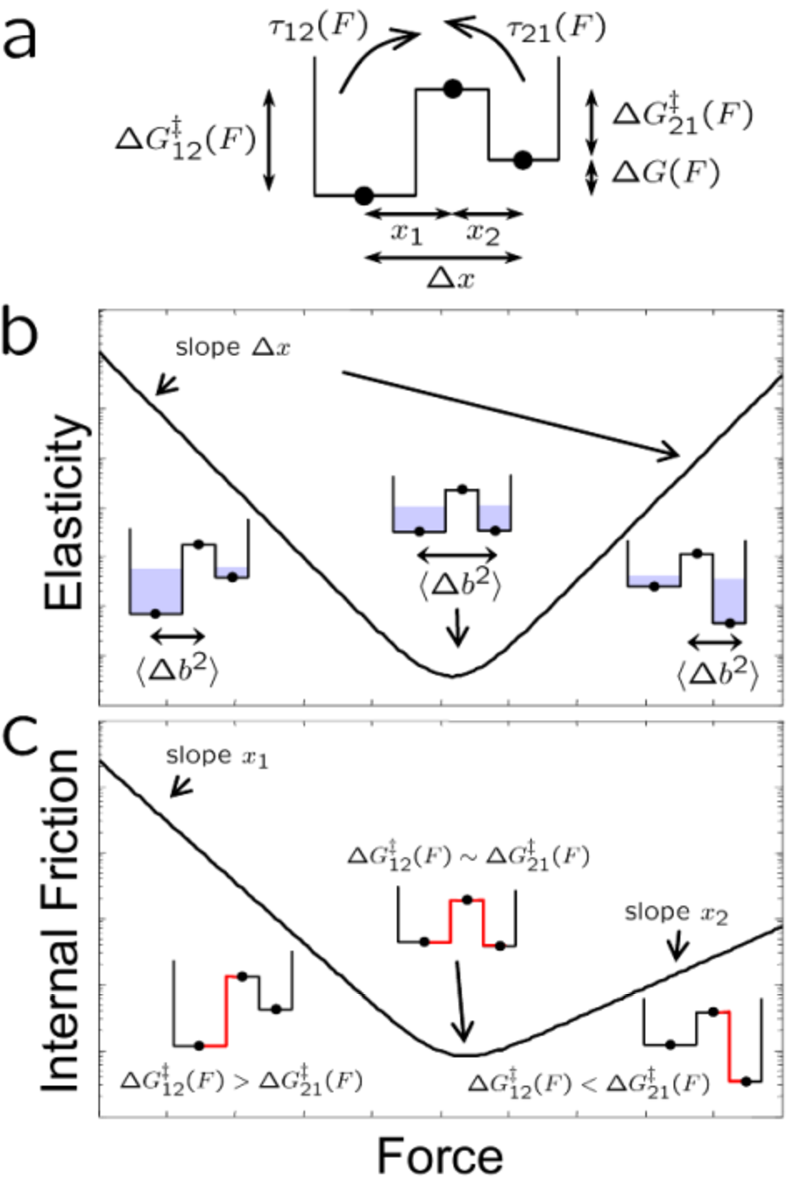}
\end{center}
\begin{center}Figure 5\end{center}

\begin{center}
\includegraphics[width=8cm]{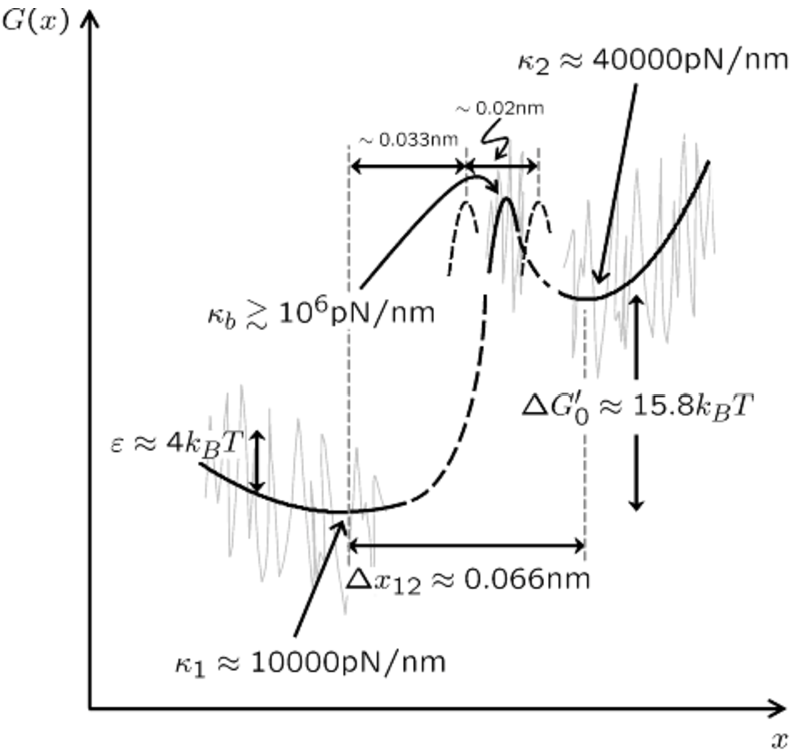}
\end{center}
\begin{center}Figure 6\end{center}

\section*{Appendices}

\setcounter{equation}{0}
\vspace{12pt}
\subsection{Dynamic compliance of parallel system
elements}

Here we derive the total dynamic compliance or response function
$J_T(\omega)$ for the cantilever and polymer in parallel, which each
have response functions $J_X(\omega)$ and $J_{\Delta R}(\omega)$,
respectively. Starting in the time domain, we can write down the
solution for the cantilever and polymer motion as:

\begin{displaymath}
\Delta R(t) = \int_0^t J_{\Delta
R}(t-t')\left(\frac{1}{2}F(t')-f(t')\right)dt'
\end{displaymath}

\begin{displaymath}
X(t) = \int_0^t J_X(t-t')\left(\frac{1}{2}F(t')+f(t')\right)dt'
\end{displaymath}

\noindent where $F$ represents an external force applied to the
system and $f$ the internal force that they share according to
Newton's $3^{rd}$ Law. By definition, the Green's response of the
whole system is its response to a unit impulse of force, so we let
$F(t)=\eta\delta(t)$, where $\delta(t)$ is the Dirac
``delta-function'' and $\eta$ the size of impulse. This gives

\begin{displaymath}
\Delta R(t) = \frac{\eta}{2}J_{\Delta R}(t) - \int_0^t J_{\Delta
R}(t-t')f(t')dt'
\end{displaymath}

\begin{displaymath}
X(t) = \frac{\eta}{2}J_X(t) + \int_0^t J_X(t-t')f(t')dt'
\end{displaymath}

\noindent Taking the Fourier Transform of these (presuming all
response functions are zero for $t<0$) we find

\begin{equation}\label{DeltaR}
\Delta R(\omega) = \frac{\eta}{2}J_{\Delta R}(\omega) - J_{\Delta
R}(\omega)f(\omega)
\end{equation}

\begin{equation}\label{DeltaX}
X(\omega) = \frac{\eta}{2}J_X(\omega) +
J_X(\omega)f(\omega)
\end{equation}

\noindent Thus, using the fact that the cantilever and polymer
displacements must be the same for all times ($\Delta R(t)=X(t)$),
we can solve for the internal force $f(\omega)$:

\begin{displaymath}
f(\omega) = \frac{\eta}{2}\frac{J_{\Delta R}(\omega) -
J_X(\omega)}{J_{\Delta R}(\omega) + J_X(\omega)}
\end{displaymath}

\noindent this can then be plugged back into Eq.(\ref{DeltaR}) or
Eq.(\ref{DeltaX}), to give the total dynamic compliance as the
displacement response due to a unit delta function input:

\begin{displaymath}
J_T(\omega) = \frac{\Delta R(\omega)}{\eta} = \frac{X(\omega)}{\eta}
= \frac{J_X(\omega)J_{\Delta R}(\omega)}{J_X(\omega)+J_{\Delta
R}(\omega)}
\end{displaymath}\\

\subsection{Frictional Freely Jointed Chain}

To model the molecular viscoelasticity of a polymer at forces which
are small (approximately, $F<500$pN), we develop a Frictional FJC
(FFJC) model of rods interconnected with joints with constant
friction and calculate the form of $\zeta_{FJC}(F)$. We focus on a
single monomer, assuming that each rod of the FJC is statistically
independent, so the stiffness of each rod will add mechanically in
series to the stiffness of the whole chain. Typical monomer/rod
lengths for polysaccharides are $b\sim 1$nm, so our experiments are
in the regime where $F\gg k_BT/b$ and the elasticity spectrum can be
calculated from statistical mechanics as

\begin{equation}\label{KappaFJC}
\kappa_{FJC}(F) = \frac{F^2}{k_BT}
\end{equation}

\noindent To model the internal friction of a FJC we again focus on
a single monomer/rod in the high force regime and consider that to
rotate such a rod there is some friction $\zeta_{\theta}$ opposing
this motion, which we presume is constant and associated with the
internal friction of `joints' between rods. The rotational equation
of motion for a segment or rod of length $b$ held under a large
tensile force $F$ ($Fb\gg k_BT$) will be

\begin{equation}\label{tau_theta}
\zeta_{\theta}\dot{\theta}(t)=-Fb\theta
\end{equation}

\noindent Now we consider how these dynamics project onto the line
of applied force. The change in projected length of the monomer
compared to its actual length will be $\Delta b=b(1-\cos\theta)$,
which in the small angle limit will be:

\begin{equation}\label{Delta_b}
\Delta b\approx\frac{1}{2}b\theta^2
\end{equation}

\noindent Differentiating $\Delta b$, and using Eq.(\ref{tau_theta})
\& Eq.(\ref{Delta_b}), we find its equation of motion to be:

\begin{displaymath}
\dot{\Delta b}= b\theta\dot{\theta}=
-\frac{Fb^2\theta^2}{\zeta_{\theta}}=
-\frac{2Fb}{\zeta_{\theta}}\Delta b
\end{displaymath}

\noindent This again has an exponentially decaying solution $\Delta
b(t)\sim e^{-t/\tau_{FJC}}$, with time constant
$\tau_{FJC}=\frac{\zeta_{\theta}}{2Fb}$. Now $\tau_{FJC} =
\zeta_{FJC}/\kappa_{FJC}$ and thus, using Eq.(\ref{KappaFJC}), the
effective friction along the $z$ direction is then

\begin{equation}\label{ZetaFJC}
\zeta_{FJC}(F) = \frac{\zeta_{\theta}}{2k_bTb}F
\end{equation}

\noindent which predicts a linear increase of the internal friction
constant with force, presuming $\zeta_{\theta}$ is constant. \\

\section*{Additional Figure}

\begin{figure}[!htb]
\begin{center}
\includegraphics[width=10cm]{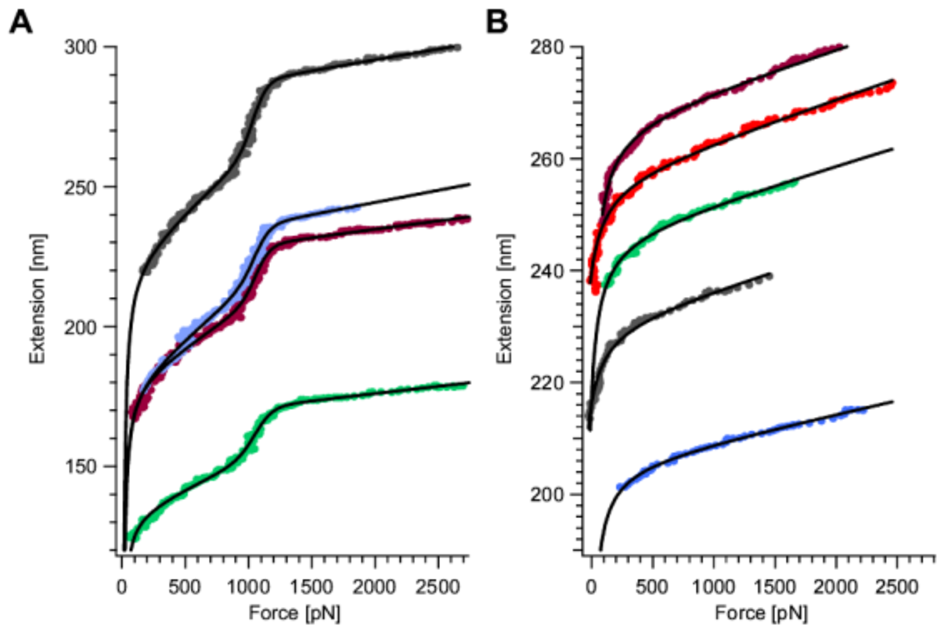}
\end{center}
\end{figure}

\noindent {\bf Fig. 7.} Extension-Force traces from constant pulling
rate experiments for (A) 4 different dextran molecules and (B) 5
different cellulose molecules. Each trace represents the final stage
(marked (3) in Fig.2 in main text) of the thermal noise force-clamp
spectroscopy protocol from which we determine the contour length
$L_c$. For cellulose we fit the extension as function of force,
using a FJC of elastic segments (FJC+), where the chain is
characterised by a number of Kuhn segments $N_k$, which have length
$b$ and elasticity $\kappa$ \cite{Smith96}, to give a contour length
$L_c=N_kb$. For dextran, we assume a Boltzmann weighted sum of the
two states, where the monomer lengths in the chair and boat state
are represented by different Kuhn segment lengths and elasticity of
a FJC+ model:$$ \langle\Delta R(F)\rangle = \frac{N_k}{1+e^{-\Delta
G(F)}}\left(b_1\mathcal{L}(Fb_1)\left(1+\frac{F}{\kappa_1b_1}\right)
+ b_2\mathcal{L}(Fb_2)\left(1+\frac{F}{\kappa_2b_2}\right)e^{-\Delta
G(F)}\right),$$ where $\mathcal{L}$ is the Langevin function, $F$ is
the imposed tension, and $\Delta G(F)= \Delta G_0- F\Delta x_{12}$,
with $\Delta x_{12}$ being the spatial separation between the two
states. Factors of $k_BT$ are dropped for clarity. For dextran, we
assume the contour length is given by $L_c\approx N_kb_1$.\\

%

\end{document}